\documentclass[letter]{ptptex}
\usepackage{latexsym}
\usepackage{amssymb}
\usepackage{amsfonts}
\usepackage{amsmath}
\usepackage{bm}
\usepackage[dvips]{graphicx}
\usepackage[usenames]{color}
\usepackage{ulem}
\usepackage{multirow}
\allowdisplaybreaks
\newcommand{\e}{\epsilon}

\newcommand{\m}{\mu}
\newcommand{\n}{\nu}

\begin{document}
%\markboth{Fujita}{14PN energy flux}
\title{Gravitational radiation for extreme mass ratio inspirals\\to the 14th post-Newtonian order
}
  % ----------------------
\author{Ryuichi Fujita$^{1,2}$
}
\inst{
${}^1$Raman Research Institute, Bangalore 560 080, India\\
${}^2$Departament de F\'isica, Universitat de les Illes Balears, 
Cra.\ Valldemossa Km.\ 7.5, Palma de Mallorca, E-07122 Spain
}
  % ----------------------

\abst{
We derive gravitational waveforms needed to compute the 14th post-Newtonian 
(14PN) order energy flux for a particle in circular orbit around a 
Schwarzschild black hole, i.e. $v^{28}$ beyond the leading Newtonian 
approximation where $v$ is the orbital velocity of a test particle.
We investigate the convergence of the energy flux in the PN 
expansion and suggest a fitting formula which can be used to extract 
unknown higher order PN coefficients from accurate numerical data for 
more general orbits around a Kerr black hole. 
The phase difference between 
the 14PN waveforms and numerical waveforms after two years inspiral 
is shown to be about $10^{-7}$ for $\mu/M=10^{-4}$ and $10^{-3}$ 
for $\mu/M=10^{-5}$ where $\mu$ is the mass of a compact object 
and $M$ the mass of the central supermassive black hole. 
In first order black hole perturbation theory, for extreme mass ratio inspirals 
which are one of the main targets of Laser Interferometer Space Antenna, 
the 14PN expressions will lead to the data analysis accuracies
comparable to the ones resulting from high precision numerical waveforms.
}
\maketitle
%%%%%%%%%%%%%%%%%%%%%%%%%%%%%%%%%%%%%%%%%%%%%%%%%%%%%%%%%%%%%%%%%%%%%%%%%%%
{\it{Introduction}}.~Gravitational waves emitted from a stellar mass compact 
object of mass $\m(\sim 1-100M_{\odot})$ orbiting a supermassive black hole 
of mass $M(\sim 10^{5}-10^{7}M_{\odot})$ at the centers of galaxies are one of 
the main 
astrophysical sources for the Laser Interferometer Space Antenna (LISA). 
The expected number of events of these extreme mass ratio inspirals (EMRIs) 
detected by LISA during its mission will be 
$10-1000$~\cite{ref:GairEtal2004} 
(see also a recent work on the capture rate by 
the 2.5PN $N$-body simulations~\cite{ref:MerrittEtal2011}). 
Observation of gravitational waves from EMRIs will provide information of 
the source, such as the masses, the spin of the black hole and the distribution 
of compact objects in the centers of galaxies. 
To extract such information by data analysis, 
the accumulated difference of phase between templates and the true signal 
during the observation should be less than one radian. 
The total number of wave cycles detected by LISA is $10^{5}-10^{6}$ since 
LISA has maximum sensitivity around $10^{-3}$Hz and its mission time is a few years. 
Thus, for LISA parameter estimation of EMRIs, 
the accuracy of theoretical waveforms should be better than $10^{-5}$. 

Since the mass ratio is extreme, EMRIs can be described by 
black hole perturbation theory, which uses the mass ratio $\m/M$ 
as an expansion parameter (see Ref.~\citen{LZ2011} for 
fully nonlinear numerical simulations of $\m/M=1/100$). 
At the lowest order, the small body traces a geodesic in the black hole geometry. 
Over time scales of order $\sim M^2/\mu$ (we use units $G=c=1$) however, 
the orbit evolves adiabatically due to the small body's 
interaction with its own gravitational field, 
i.e. gravitational self-force~\cite{PPV2011,ref:Thornburg2011}. 
This means that the gravitational self-force has to be 
taken into account for LISA parameter estimation of EMRIs. 
Second order black hole perturbation theory may also be required for 
the parameter estimation and is of great interest when the mass ratio is large. 

Although black hole perturbation theory is powerful enough 
that one can compute 
gravitational waves accurately even in the strong field, 
the computational cost is very high to perform calculations which cover 
all the parameter space of EMRIs~\cite{ref:GairEtal2004}. 
This motivates us to use analytic modeling of gravitational waves by 
the post-Newtonian (PN) theory, which uses the orbital velocity of a compact object 
$v=\sqrt{M/r_0}$, 
where $r_0$ is the orbital radius, 
as an expansion parameter. 

In this letter, we investigate the extent of which PN order waveforms 
improve the solution for LISA parameter estimation of EMRIs 
by using the first order black hole perturbation theory. 
Currently, the highest available PN order for gravitational waveforms for EMRIs is 5.5PN 
($v^{11}$ beyond Newtonian order) for a test particle in a circular orbit 
around a Schwarzschild black hole~\cite{ref:FI2010}. 
However, 
the convergence of the PN expansion becomes worse when $v$ is larger and 
the phase difference between the PN and numerical waveforms exceeds 
one radian after two years inspiral~\cite{ref:EOB_EMRI, ref:FI2010}. 
In this work, we improve on the accuracy of the energy flux to 14PN and 
exhibit clearly its closeness to a high precision numerical computation of the energy flux. 
We would like to point out
that such high PN order computations have not been performed up till now, 
since the number of terms necessary to derive the PN waveforms 
grows exponentially when the PN order becomes higher. For instance,
our current code uses $70$, $3.3\times 10^{2}$, $9.0\times 10^{2}$, 
$1.9\times 10^{3}$, $5.6\times 10^{3}$ and $1.1\times 10^{4}$MBytes memory, 
taking seconds to half an hour, to compute multipolar waveforms for 
$\ell=m=2$ mode at 6PN, 10PN, 12PN, 14PN, 16PN and 18PN respectively. 
Thus, it will be difficult to obtain 19PN or higher order expressions 
with reasonable time on a personal computer by using our current code. 
We also suggest the fitting formula of the energy flux which 
can be used for more general orbits around a Kerr black hole. 
One of the important consequence of this work for LISA is our demonstration
that the phase difference after two years inspiral 
between our new 14PN waveforms and high precision numerical waveforms 
is negligible. 
This indicates that the 14PN expressions will lead to accuracies in
LISA data analysis for EMRIs over two years comparable to accuracies 
resulting from high accuracy numerical waveforms. 

%%%%%%%%%%%%%%%%%%%%%%%%%%%%%%%%%%%%%%%%%%%%%%%%%%%%%%%%%%%%%%%%%%%%%%%%%%%
{\it{The method}}.~The fundamental equation of black hole perturbation is 
the Teukolsky equation, which describes 
the gravitational perturbation of a black hole in terms of 
the Weyl scalar $\Psi_4$~\cite{Teukolsky:1973ha}. 
$\Psi_4$ is related to the gravitational wave polarizations at infinity as
$\Psi_4\rightarrow (\ddot{h}_{+}-i\,\ddot{h}_{\times })/2$ 
where dot, $\dot{\,}$, denotes time derivative, $d/dt$. 
The Teukolsky equation can be separated 
if we expand $\Psi_4$ in the Fourier domain using 
the $-2$ spin-weighted spheroidal harmonics~\cite{Teukolsky:1973ha}. 
For the case of Schwarzschild black hole, 
the radial Teukolsky equation reads 
\begin{equation}
 \left[\Delta^2{d\over dr}\left({1\over \Delta}{d\over dr}\right)+U(r)\right]
  R_{\ell m\omega}(r) = T_{\ell m\omega}(r),
\label{eq:Teu}
\end{equation}
where 
$\Delta=r(r-2M)$, $U(r)$ is the potential 
and $T_{\ell m\omega}$ is the source term depending on the particle's orbit. 

The radial Teukolsky equation can be solved by the Green function method. 
In this work, we use a formalism developed by Mano, Suzuki and Takasugi (MST) 
to obtain homogeneous solutions of the Teukolsky equation~\cite{ref:MST_R}. 
In this formalism, the homogeneous solutions of the Teukolsky equation are
expressed in a series of hypergeometric functions and Coulomb wave functions, 
which converge at the horizon and infinity respectively. 
The formalism is very powerful for the PN expansion of 
the Teukolsky equation since the series expansion is closely related to 
the low frequency expansion. The expansion coefficients of the two series $a_{n}^{\n}$ 
are the same and satisfy the three-term recurrence relation
\begin{eqnarray}
\alpha_n^\nu a_{n+1}^{\n}+\beta_n^{\nu} a_{n}^{\n}+\gamma_n^\nu a_{n-1}^{\n}=0,
\label{eq:3term}
\end{eqnarray}
where $\nu=\ell+O(\omega^2)$, $\alpha_n^\nu=O(\omega)$, $\beta_n^{\nu}=O(1)$ and 
$\gamma_n^\nu=O(\omega)$. 
We note that the parameter $\nu$ does not exist in the original Teukolsky equation and 
is determined so that the series converges and 
represents a solution of the Teukolsky equation. 
One can derive the low frequency expansion of $a_{n}^{\n}$ by solving 
Eq.~(\ref{eq:3term}) iteratively. 
Thus, if we derive $a_{n}^{\n}$ up to a required order, 
we automatically obtain the PN expansion formulas up to the required order. 
See Ref.~\citen{ST} for details of the formalism. 

Using the MST formalism, the energy flux absorbed into the horizon was derived 
up to relative 4PN (i.e. 6.5PN beyond Newtonian order) 
for a test particle in a circular and equatorial orbit around 
a Kerr black hole~\cite{TMT}. 
Gravitational wave flux to infinity was also 
computed up to 2.5PN for a test particle in slightly eccentric and inclined 
orbits around a Kerr black hole~\cite{STHGN,Ganz}. 
More recently, the MST formalism was applied to obtain 
the 5.5PN waveforms for a test particle in a circular orbit 
around a Schwarzschild black hole~\cite{ref:FI2010} and the 4PN waveforms 
for a test particle in a circular and equatorial orbit 
around a Kerr black hole~\cite{ref:spin_resum}, 
confirming the 5.5PN energy flux in Ref.~\citen{ref:TTS} and 
the 4PN energy flux in Ref.~\citen{ref:TSTS} respectively. 

%%%%%%%%%%%%%%%%%%%%%%%%%%%%%%%%%%%%%%%%%%%%%%%%%%%%%%%%%%%%%%%%%%%%%%%%%%%
{\it{Results and Analysis}}.~Once we compute the homogeneous solutions of 
the Teukolsky equation, using the Green function method 
we build the solution of the Teukolsky equation Eq.~(\ref{eq:Teu}), which is 
purely outgoing at infinity and ingoing at the horizon. 
For the case of a test particle in a circular orbit around a Schwarzschild black hole, 
the gravitational wave flux to infinity is given by 
\begin{equation}
{dE\over dt}=\sum_{\ell=2}^{\infty} \sum_{m=-\ell}^{\ell}
   \frac{\vert Z_{\ell m\omega}\vert^2}{4\pi \omega^2},
\label{eq:flux}
\end{equation}
where $\omega =m\Omega$ is the frequency of gravitational waves, 
$\Omega=\sqrt{M/r_0^3}$ is the angular frequency of the orbit and 
$Z_{\ell m\omega}$ is derived from the asymptotic behavior of 
the solution of the Teukolsky equation at infinity. 

We compute gravitational waveforms which are 
necessary for computing the 14PN energy flux to infinity 
for a test particle in a circular orbit around a Schwarzschild black hole. 
We show only the next new 6PN terms of the energy flux, $dE^{(12)}/dt$, 
because of space restrictions. 
\begin{align}
{dE^{(12)}\over dt}=&
{\frac {2067586193789233570693}{602387400044430000}}-{\frac {246137536815857}{157329572400}}\,\gamma-{\frac {271272899815409}{157329572400}}\,\ln  \left( 2 \right) 
\cr
&
-{\frac {246137536815857}{157329572400}}\,\ln  \left( v \right)
 -{\frac {27392}{105}}\,\zeta  \left( 3 \right) -{\frac {54784}{315}}\,\ln  \left( 2 \right) {\pi }^{2}-{\frac {27392}{315}}\,{\pi }^{2}\ln  \left( v \right) 
\cr
&
+{\frac {5861888}{11025}}\,\ln  \left( 2 \right) \ln  \left( v \right) +{\frac {3803225263}{10478160}}\,{\pi }^{2}-{\frac {256}{45}}\,{\pi }^{4}
+{\frac {5861888}{11025}}\, \left( \ln  \left( 2 \right)  \right) ^{2}
\cr
&
+{\frac {1465472}{11025}}\, \left( \ln  \left( v \right)  \right) ^{2}-{\frac {27392}{315}}\,{\pi }^{2}\gamma+{\frac {2930944}{11025}}\,\ln  \left( v \right) \gamma+{\frac {5861888}{11025}}\,\ln  \left( 2 \right) \gamma
\cr
&
+{\frac {1465472}{11025}}\,{\gamma}^{2}
-{\frac {437114506833}{789268480}}\,\ln  \left( 3 \right) -{\frac {37744140625}{260941824}}\,\ln  \left( 5 \right),
\end{align}
where $\zeta(n)$ is the Zeta function. 
We note that $(\ln (v))^2$ term appears at 6PN. 
We also note that $(\ln (v))^3$ and $(\ln (v))^4$ terms appear 
from 9PN and 12PN respectively. 
One of the reasons is that $(\ln (v))^n$ terms are produced by the PN expansion of 
$z^{\nu}$ in the homogeneous solution of the Teukolsky equation, where 
$z=\omega r$ and $\nu=\ell+O(v^6)$. 
These $(\ln (v))^n$ terms agree with Eq.~(44) in Ref.~\citen{ref:GR2010}, which 
predicts $(\ln (v))^n$ terms at 3$n$-PN using the renormalization group equations. Our present work suggests that to obtain unknown PN terms by 
fits to numerical results one must
include $(\ln (v))^n$ terms starting from 3$n$-PN with the structure 
\begin{equation}
{dE\over dt}=\sum_{k=0}^{\infty}\sum_{p=0}^{[k/6]}{dE^{(k,p)}\over dt}(\ln (v))^p\,v^k, 
\label{eq:flux_fit}
\end{equation}
where $[\cdots]$ is the floor function. 
Note that one can use the fitting formula Eq.~(\ref{eq:flux_fit}) 
for more general orbits around a Kerr black hole, 
which are known at most up to 4PN~\cite{ref:TSTS,Ganz}. 

We use the factorized multipolar waveforms 
introduced in Ref.~\citen{DIN} to compute the gravitational waveforms. 
The multipolar waveforms are decomposed 
into five factors as 
\begin{eqnarray}
h_{\ell m}=h_{\ell m}^{({\rm N},\e_p)}\,\hat{S}_{\rm eff}^{(\e_p)}\,T_{\ell m}\,e^{i\delta_{\ell m}}(\rho_{\ell m})^\ell\,, 
\end{eqnarray}
where $h_{\ell m}^{({\rm N},\e_p)}$ is the Newtonian contribution to waveforms 
and $\e_p$ denotes the parity of the multipolar waveforms. 
In the case of circular orbits, $\e_p=0$ ($\e_p=1$) when $\ell+m$ is even (odd). 
The $\ell$-th root of the amplitude $\rho_{\ell m}$ improves the convergence of 
the PN expansion by dealing with the linear term of $\ell$ in the 1PN terms 
of the amplitudes. See Ref.~\citen{DIN} for the definitions of the other terms. 

The factorized multipolar waveform is the simplest and the most efficient 
resummation technique so far. 
The energy flux computed from the factorized resummed waveforms agrees better 
with the numerical results than the ones from Taylor-expanded waveforms 
and other resummed waveforms as Pad\'e approximation~\cite{DIN}. 

In this letter, we show only the next new 6PN terms of $\rho_{2,2}$, which was
earlier computed 
up to 5.5PN~\cite{DIN,ref:FI2010}. The complete expressions will be shown 
elsewhere~\cite{ref:rho14pn_flux}. 

\begin{align}
\rho _{2, \,2}^{(12)}=&
{\frac {313425353036319023287}{1132319812111488000}}
-{\frac {6848}{105}}\,\zeta  \left( 3 \right) 
-{\frac {91592}{11025}}\,{\pi }^{2} 
\cr
&
-{\frac {241777319107}{3208936500}}\,{\rm eulerlog} \left( 2,v \right) 
+{\frac {91592}{11025}}\, \left( {\rm eulerlog} \left( 2,v \right)  \right) ^{2},
\end{align}
where $\rho _{2, \,2}^{(n)}$ is $O(v^{n})$ coefficient of $\rho _{2, \,2}$, 
${\rm eulerlog}(2, \,v) \equiv \gamma + \ln(4v)$ and $\gamma$ is the Euler constant. 
Again, we note that $(\ln (v))^2$ term appears at 6PN. 

\begin{figure}[t]
\begin{center}
\includegraphics[width=70mm]{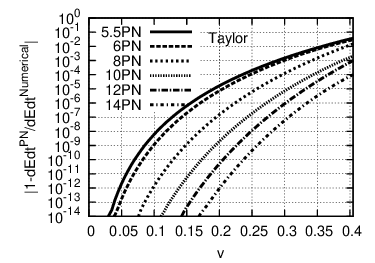}%
\includegraphics[width=70mm]{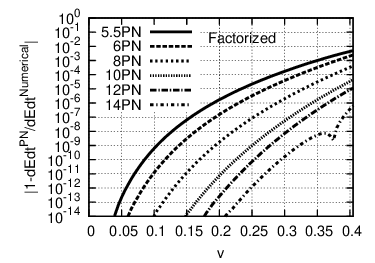}
\end{center}
\caption{\label{fig:flux}Absolute values of the difference of energy flux 
to infinity between numerical results and 
Taylor-expanded (left) or factorized (right) PN approximation 
as a function of orbital velocity. Note that 14PN flux 
converges very well even around ISCO, $v=1/\sqrt{6}=0.40825$. 
}
\end{figure}

Fig.~\ref{fig:flux} shows the comparison of energy flux to infinity 
between the PN approximations and the numerical calculation. 
To demonstrate the efficiency of the factorized waveforms, 
we show the results of the PN energy flux using 
the Taylor-expanded waveforms (left panel) and 
the factorized waveforms (right panel) in Fig.~\ref{fig:flux}. 
The numerical energy flux is obtained 
using the high precision code in Ref.~\citen{FT1_2}. The accuracy of 
the numerical calculation is 
better than $10^{-10}$ if we set the maximum value of $\ell$ to $20$ 
in Eq.~(\ref{eq:flux}). The $n$-PN flux needs $\ell$ up to $n+2$. 
The agreement of the total energy flux between 
the PN and the numerical results becomes better 
when the PN order is higher even around 
the innermost stable circular orbit (ISCO) 
(see Ref.~\citen{OT2000} for the calculation beyond ISCO). 

%%%%%%%%%%%%%%%%%%%
%%%%%%%%%%%%%%%%%%%
To investigate quantitatively the PN order needed 
for LISA parameter estimation of EMRI, we compare the phase difference during two years 
quasi-circular inspiral between the factorized PN waveforms and the numerical waveforms. 
Following Ref.~\citen{ref:EOB_EMRI}, we examine two systems, which are named 
System-I and System-II. 
System-I has masses $(M,\m)=(10^5,10)M_{\odot}$ 
and inspirals from $r_0\simeq 29.34M$ to $r_0\simeq 16.1M$, 
whose frequency sweeps from $f_{\rm GW}\simeq 4\times 10^{-3}$Hz to 
$f_{\rm GW}\simeq 10^{-2}$Hz. 
System-II has masses $(M,\m)=(10^6,10)M_{\odot}$ 
and starts inspiral from $r_0\simeq 10.6M$ to $r_0\simeq 6.0M$, 
whose frequency sweeps from $f_{\rm GW}\simeq 1.8\times 10^{-3}$Hz to 
$f_{\rm GW}\simeq 4.4\times 10^{-3}$Hz. 
System-I (II) has $\sim 1\times 10^{6}$ ($\sim 5\times 10^{5}$) rads 
of orbital phase during its inspiral and corresponds to the early (late) inspiral phase 
of an EMRI in the frequency band of LISA. 

\begin{figure}[t]
\begin{center}
\includegraphics[width=70mm]{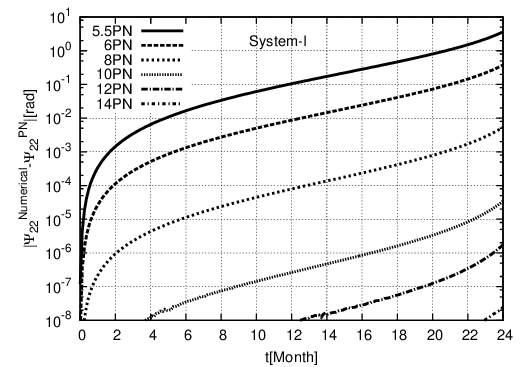}%
\includegraphics[width=70mm]{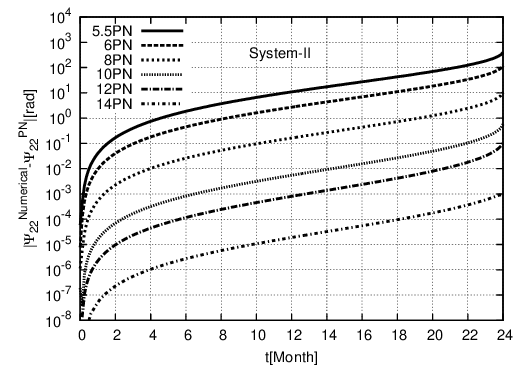}
\caption{Absolute values of the phase difference due to two years inspiral 
between the factorized PN and 
the numerical waveforms for $h_{2,2}$ as a function of time in month. 
The left panel shows the dephase 
for $(M,\m)=(10^5,10)M_{\odot}$, which starts from 
$r_0\simeq 29.34M$ to $r_0\simeq 16.1M$ and sweeps frequencies 
$f_{\rm GW}\in [4\times 10^{-3},10^{-2}]$Hz. 
The right panel shows the dephase 
for $(M,\m)=(10^6,10)M_{\odot}$ 
from $r_0\simeq 10.6M$ to $r_0\simeq 6.0M$ 
with associated frequencies $f_{\rm GW}\in [1.8\times 10^{-3},4.4\times 10^{-3}]$Hz. 
The left (right) panel represents the early (late) inspiral phase in the LISA band.
If the dephase is less than 10 milliradians, 
the PN waveforms will provide the parameter estimation of EMRI comparable to 
the one resulting from numerical waveforms. 
}\label{fig:dephase22}
\end{center}
\end{figure}

Fig.~\ref{fig:dephase22} shows the comparison of the phase between 
the dominant mode $h_{2,2}$ of the factorized PN and the numerical calculation 
(see Ref.~\citen{Hughes2001} for the calculation of the phase). 
For System-I (II), the absolute values of the phase difference 
between the factorized PN waveforms and the numerical waveforms 
after two years inspiral are about $3.6$ ($4.0\times 10^{2}$), 
$0.38$ ($1.2\times 10^{2}$), $5.4\times 10^{-3}$ ($10$), 
$3.4\times 10^{-5}$ ($0.57$), $1.8\times 10^{-6}$ ($0.11$) and 
$2.5\times 10^{-8}$ ($8.9\times 10^{-4}$) rads 
for 5.5PN, 6PN, 8PN, 10PN, 12PN and 14PN respectively 
(The relative error of the amplitude between the factorized 14PN and 
the numerical waveforms is $6.6\times 10^{-13}$ ($7.1\times 10^{-7}$) 
for System-I (II)).

The strongest EMRI events detected by LISA may have the signal to noise ratio 
up to $\rho\sim 100$~\cite{ref:IMRI_EMRI2007}. 
Thus, LISA can measure phase difference of the order of 
$1/\rho\sim 10$ milliradians by matched filtering~\cite{ref:Thornburg2011}. 
This suggests that the 8PN (14PN) waveforms will lead to LISA 
parameter estimation of System-I (II) comparable to the one using 
numerical waveforms. 
Since System-II represents the inspiral in the most strong-field 
of a Schwarzschild black hole, 
the 14PN waveforms will provide LISA parameter estimation of EMRI 
comparable to the one resulting from numerical waveforms. 

%%%%%%%%%%%%%%%%%%%
%%%%%%%%%%%%%%%%%%%%%%%%%%%%%%%%%%%%%%%%%%%%%%%%%%%%%%%%%%%%%%%%%%%%%%%%%%%
{\it{Conclusion and Discussion}}.~Using the first order black hole perturbation theory, 
we have computed 
gravitational waveforms consistent with the 14PN energy flux for a test particle 
in a circular orbit around a Schwarzschild black hole. 
The high PN order computation has been 
performed systematically using the MST formalism~\cite{ref:MST_R}. 
We provide a fitting formula of the energy flux, Eq.~(\ref{eq:flux_fit}), 
for more general orbits around a Kerr black hole. 
Comparing the energy flux with a high precision numerical computation, 
we investigated the approach of the PN expansion towards the numerical results
 and 
found that the PN expansion converges well even at ISCO. 
The phase difference between 
our new analytic waveforms and the numerical waveforms after 
two years of inspiral is so negligibly small that it does not 
lead to any discrepancy in LISA data analysis.
Thus, using first order black hole perturbation theory, 
one can derive sufficiently high PN order waveforms analytically and build
the template banks to span the parameter space efficiently without recourse 
to a more expensive numerical computation. 

The EMRIs contain more generic systems such as eccentric and inclined 
orbits of a test particle around a Kerr black hole 
(see Ref.~\citen{ref:TMSS1996} for an application to a spinning particle). 
The orbital velocity at ISCO can be larger when the black hole is rotating. 
If the orbits become more generic, one needs a larger number of wave modes, 
which will contain higher frequency contributions than 
the case in this work. 
Thus, for more generic orbits around a Kerr black hole 
we have to compute higher PN orders 
than 14PN in order to achieve good agreement with the numerical waveforms. 
Since the formalism adopted in this work is systematic and does not have 
theoretical issues to compute higher PN order, in principle one can compute 
sufficiently high PN order waveforms even for generic orbits. 
It will be possible to perform sufficiently high PN 
order computation for the case of a test particle in slightly eccentric and 
inclined orbits around a Kerr black hole. 
However, it may be difficult to compute them for the case of a test particle in 
large eccentric and inclined orbits around a Kerr black hole in reasonable time. 
Moreover, conservative effects of self-force, which are not included 
in this work, may contribute $\sim 20$ radians to the phase of waveforms 
during inspiral~\cite{ref:HG2009} (however see Table ~II in 
Ref.~\citen{ref:HJNSST}, which suggests that conservative effects 
up to 9PN is sufficient to reduce the phase error to 10 milliradians 
by investigating 18PN scalar self-force). 
Another approach to discuss generic orbits may be 
the effective-one-body formalism which 
can determine unknown terms in PN approximation and self-force effects 
by calibrating them with numerical calculation~\cite{ref:EOB_EMRI,ref:NBHPBMT}. 
The present work has implications for this approach too.
For the calibration, we recommend the use of our proposed
 fitting formula Eq.~(\ref{eq:flux_fit}) 
dealing with $(\ln (v))^n$ terms appearing from 3$n$-PN. 

%%%%%%%%%%%%%%%%%%%%%%%%%%%%%%%%%%%%%%%%%%%%%%%%%%%%%%%%%%%%%%%%%%%%%%%%%%%
{\it{Acknowledgments}}.~We are grateful to Bala R. Iyer for continuous 
encouragement to look into this problem and useful comments on the 
manuscript. We also thank Hiroyuki Nakano for useful comments on the 
manuscript. 
%%%%%%%%%%%%%%%%%%%%%%%%%%%%%%%%%%%%%%%%%%%%%%%%%%%%%%%%%%%%%%%%%%%%%%%%%%%

%%%%%%%%%%%%%%%%%%%%%%%%%%%%%%%%%%%%%%%%%%%%%%%%%%%%%%%%%%%%%%%%%%%%%%%%%%%
\end{document}